\documentclass[aps,groupedaddress, preprint]{revtex4}
\usepackage{graphicx,epsf,amssymb,amsmath,latexsym,epsfig}
\newcommand\pa{\partial}

\newcommand\be{\begin{equation}}
\newcommand\ee{\end{equation}}

\newcommand\lab[1]{\label{eq:#1}}

\begin{document}

\title{Photon Production From The Scattering of Axions Out of a Solenoidal Magnetic Field}

\author{Eduardo I. Guendelman}
\email{guendel@bgu.ac.il}
\author{Idan Shilon}
\email{silon@bgu.ac.il}

\affiliation{Physics Department, Ben-Gurion University of the Negev, Beer-Sheva 84105, Israel}

\author{Giovanni Cantatore}
\email{cantatore@trieste.infn.it}
\affiliation{Universit\'a and INFN Trieste, via valerio 2, 34127 Trieste, Italy}

\author{Konstantin Zioutas}
\email{Konstantin.Zioutas@cern.ch}
\affiliation{University of Patras, Patras, Greece}

\begin{abstract}
We calculate the total cross section for the production of photons from the scattering of axions by a strong inhomogeneous magnetic field in the form of a 2D $\delta$-function, a cylindrical step function and a 2D Gaussian distribution, which can be approximately produced by a solenoidal current. The theoretical result is used to estimate the axion-photon conversion probability which could be expected in a reasonable experimental situation. The calculated conversion probabilities for QCD inspired axions are bigger by a factor of 2.67 (for the cylindrical step function case) than those derived by applying the celebrated 1D calculation of the (inverse) coherent Primakoff effect. We also consider scattering at a resonance $E_{axion} \sim m_{axion}$, which corresponds to the scattering from a $\delta$-function and gives the most enhanced results. Finally, we analyze the results of this work in the astrophysical extension to suggest a way in which they may be directed to a solution to some basic solar physics problems and, in particular, the coronal heating problem.
\end{abstract}

\maketitle

\setcounter{equation}{0}

\section{Introduction}

The possible existence of a light pseudoscalar particle is a very interesting possibility. For example,
the axion \cite{Peccei} - \cite{Wilczek},  which was introduced in order to solve the strong CP problem in QCD, has since then also been postulated as a candidate for the dark matter in the universe. A great number of ideas and experiments for the direct detection of this particle have been proposed in the past \cite{Goldman}, \cite{Review}. For example, It was recognized by Sikivie that axion detection exploiting axion to photon conversion in a magnetic Þeld was a possibility \cite{sik}.

Related to that, in a series of recent publications by one of us \cite{duality}, it was shown that an axion-photon system displays a continuous axion-photon duality symmetry when an external magnetic field is present and when the axion mass is neglected. This allows one to analyze the behavior of axions and photons in external magnetic fields in terms of an axion-photon complex field. For example, the deflection of light from magnetars has been recently studied using these techniques \cite{doron}. It is important to note here that the same duality symmetry exists also when considering massive photons, under the condition $m_{\gamma} = m_{a}$, that is the photon and the axion masses are equal. These conditions can be achieved when conducting experiments where the axion-photon conversion region is filled with a suitable refractive gas. In this letter we show that the coupling of axion-photon complex particles to a localized magnetic flux generated by a solenoid renders scattering solutions with a cross section which could conceivably be measured.

To see this, let us write the Lagrangian describing the relevant light pseudoscalar coupling to the photon,

\begin{equation}
\label{axion photon ac }
\begin{array}{c}
	\mathcal{L} =  
	 -\frac{1}{4}F^{\mu\nu}F_{\mu\nu} + \frac{1}{2}\partial_{\mu}\phi \partial^{\mu}\phi - 
	\frac{1}{2}m_{a}^{2}\phi^{2} -  \vspace{4pt}
	- \frac{g}{8} \phi \epsilon^{\mu \nu \alpha \beta}F_{\mu \nu} F_{\alpha \beta} ~.
\end{array}
\end{equation}

\noindent Following Ref. \cite{idan} (and references therein), we focus on the case where an electromagnetic field with propagation along the $x$ and $y$ directions and a strong magnetic field pointing in the $z$-direction are present. The magnetic field may have an arbitrary space dependence in $x$ and $y$, but it is assumed to be time independent.

For small electromagnetic perturbations around the static magnetic background (i.e, the axion and the electromagnetic wave), we consider only small quadratic terms in the Lagrangian for the axion and the electromagnetic fields. By choosing a static magnetic field pointing in the $z$ direction and having an arbitrary $x$ and $y$ dependence and specializing to $x$ and $y$ dependent electromagnetic field perturbations and axion fields, the interaction between the background magnetic field and the axion and photon fields reduces to
 
\begin{equation}
\label{axion photon int }
	\mathcal{L}_I =  - \beta \phi E_z ~,
\end{equation}

\noindent where $\beta(x,y) = gB(x,y)$. Choosing the temporal gauge for the electromagnetic field and considering only the $z$-polarization for the electromagnetic waves (since only this polarization couples to the axion) we get the following 2+1 dimensional effective Lagrangian

\begin{equation}
\label{2 action}
	\mathcal{L}_{2} =  
	 \frac{1}{2}\partial_{\mu}A \pa^{\mu}A+ \frac{1}{2}\partial_{\mu}\phi \pa^{\mu}\phi - 
	\frac{1}{2}m_{a}^{2}\phi^{2} + \beta \phi \pa_{t} A ~,
\end{equation}

\noindent where $A$ is the $z$-polarization of the photon, so that $E_z = -\partial_{t}A$.  

Without assuming any particular $x$ and $y$ dependence for $\beta$, but insisting that 
it will be static, we see that neglecting the axion mass $m_{a}$ (the validity of this assumption will be discussed at the end of this work), we discover a continuous axion photon duality symmetry. This is due to a rotational $O(2)$ symmetry in the axion-photon field space, allowed by the axion and photon kinetic terms and by expressing the interaction term, $\mathcal{L}_I$, in an $O(2)$ symmetric way by dropping a total time derivative from it:

\begin{equation}
\label{axion photon int2}
	\mathcal{L}_I =
	\frac{1}{2}\beta(\phi \pa_{t} A - A \pa_{t}\phi) ~.
\end{equation}

Defining now the axion-photon complex field, $\Psi$, as

\begin{equation}
\label{axion photon complex}
	\Psi = \frac{1}{\sqrt{2}}(\phi + iA) 
\end{equation}

\noindent and plugging this into the Lagrangian results in

\begin{equation}
\label{ }
	\mathcal{L} = \partial_{\mu}\Psi^{*}\partial^{\mu}\Psi -\frac{i}{2}\beta(\Psi^{*}\partial_{t}\Psi - \Psi\partial_{t}	\Psi^{*}) ~,
\end{equation}

\noindent where $\Psi^{*}$ is the charge conjugation of $\Psi$. From this we obtain the equation of motion for $\Psi$

\begin{equation}
\label{equation7}
	\partial_{\mu}\partial^{\mu}\Psi + i\beta\partial_{t}\Psi = 0 ~.
\end{equation}

\noindent We therefore have the magnetic field, or $\beta/2$ (the $U(1)$ charge), coupled to a charge density. Introducing the charge conjugation \cite{part-antipart} , that is

\be
\Psi \rightarrow \Psi^{*}~,
\label{charge conjugation}
\ee

\noindent shows that the free part of the action is indeed invariant under (\ref{charge conjugation}). When acting on the free vacuum the $A$ and $\phi$ fields give rise to a photon and an axion respectively,
but in terms of the particles and antiparticles (defined in terms of  $\Psi$), we see that a photon is an antisymmetric combination of particle and antiparticle and an axion a symmetric combination, since

\be
\phi =\frac{1}{\sqrt{2}}(\Psi^{*} +\Psi) ~~\mbox{and} ~~A= \frac{1}{i\sqrt{2}}(\Psi - \Psi^{*})~.
\lab{part, antipart}
\ee

\noindent Hence, the axion is even under charge conjugation, while the photon is odd.
These two eigenstates of charge conjugation will propagate without mixing as long as no external magnetic field in the perpendicular direction to the eigenstates (i.e axion and photon) spatial dependence is applied.
The interaction with the external magnetic field is not invariant under (\ref{charge conjugation}). In fact,
under (\ref{charge conjugation}) we can see that

\be
S_I \rightarrow - S_I~,
\lab{non invariance}
\ee

\noindent where $S_{I} = \int\mathcal{L}_{I}dxdydt$. Therefore, these symmetric and antisymmetric combinations, corresponding to axion and photon, will not be preserved in the presence of $B$  in the analog QED language, since the "analog external electric potential" breaks the symmetry between particle and antiparticle and therefore will not keep in time the symmetric or antisymmetric combinations. In fact, if the corresponding external electric potential is taken to be a repulsive potential for particles, it will be an attractive potential for antiparticles, so the symmetry breaking is maximal.

Even at the classical level these two components suffer opposite forces, thus under the influence of an inhomogeneous magnetic field both a photon or an axion will be decomposed through scattering into their particle and antiparticle components, each of which is scattered in a different direction, since the corresponding electric force is related to the gradient of the effective electric potential, i.e., the gradient of the magnetic field, times the $U(1)$ charge which is opposite for particles and antiparticles. If we look at the scattering amplitudes for particles and antiparticles, we see that they have opposite signs. Calling $S$ the scattering amplitude for a particle, the amplitude for an antiparticle is then $-S$. Therefore, an axion [i.e. the symmetric combination of particle antiparticle $(1,1)$] goes under scattering to $(1,1) + (S, -S)$. So the amplitude for axion going into photon $(1,-1)$ is $S$. Hence, we conclude that the amplitude for axion-photon conversion is equal to the particle scattering amplitude.

For this effect to have meaning, we have to work at least in a 2+1 formalism \cite{splt}. The 1+1 reduction \cite{duality}, \cite{part-antipart} which allows motion only in a single spatial direction, is unable to produce such separation, since in order to separate particle and antiparticle components we need at least two dimensions
to obtain a final state with particles and antiparticles propagating in slightly different directions. 

This is in a way similar to the Stern-Gerlach experiment in atomic physics
\cite{Stern Gerlach}, where different spin orientations suffer a different deflection force proportional to the gradient of the magnetic field in the direction of the spin. Here, instead of spin we have that the photon is a combination of two states with different $U(1)$ charge and each of these components will suffer opposite force under the influence of the external inhomogeneous magnetic field. Notice also that since particle and antiparticles are distinguishable, there are no interference effect between the two processes.

Therefore an original beam of photons will be decomposed through scattering into two different elementary particle and antiparticle components (and also, of course, the photons that were not scattered). These two beams are observable, since they both have photon components, so the observable
consequence of the axion-photon coupling will be the splitting of a photon, or axion, beam by a magnetic field of the configuration considered here, whereas in the normal Primakoff effect analysis there is no explicit recognition of a splitting. This effect is, moreover, of first order in the axion-photon coupling ($g$), unlike the ``light shining through a wall phenomena'' which depend on the coupling constant squared ($g^{2}$). 

\section{first approximation: magnetic field of an infinitely thin solenoid}
\label{delta}

To apply the results of the previous section to some specific system with magnetic field, we write separately the time and space dependence of the axion-photon field as $\Psi(\vec{r},t) = \mbox{e}^{-i\omega t}\psi(\vec{r})$. 

As a first model, we are considering an inhomogeneous magnetic field of the form $B=\Phi \delta^{2}(x,y)$. This kind of a potential can not, of course, be realized in the lab, however, we will show that the results for this, presumably purely theoretical, calculation have physical significance in the resonance case, where the scattering becomes isotropic. 

Separating the time and space dependence of $\Psi$ and considering the $\delta$ function potential reduces Eq. (\ref{equation7}) to

\begin{equation}
\label{ }
	[-\vec{\nabla}^{2} + g\Phi E \delta^{2}(x,y)]\psi(\vec{r}) = E^{2}\psi(\vec{r})~.
\end{equation}

In terms of momentum space wave functions, $\phi(\vec{k}) = \int  \mbox{e}^{i\vec{k}\cdot\vec{r}}\psi(\vec{r})d^{2}r$, the latter equation is now

\begin{equation}
\label{ }
	\vec{k}^{2}\phi(\vec{k}) + g\Phi E \psi(0) = E^{2}\phi(\vec{k}) ~,
\end{equation}

\noindent from which the solution

\begin{equation}
\label{fourier}
	\phi(\vec{k}) = (2\pi)^{2}\delta^{2}(\vec{k}-\vec{k}_{0}) - \frac{g\Phi E\psi(0)}{k^{2} - E^{2}} ~,
\end{equation}

\noindent with $k_{0}^{2} = E^{2}$, is obtained. The constant $g\Phi E\psi(0)$ is determined from Eq. (\ref{fourier}) by integration over momentum space

\begin{equation}
\label{11}
	\psi(0) = 1 - g\Phi E I_{2}(-E^{2} - i\epsilon)\psi(0)~,
\end{equation}

\noindent where 

\begin{equation}
\label{ }
 I_{2}(-E^{2} - i\epsilon) = \int \frac{d^{2}k}{(2\pi)^{2}}\frac{1}{k^{2} - E^{2} -i\epsilon} = \frac{1}{4\pi}\mbox{log}(\frac{\Lambda^{2}}{z})~,
\end{equation}

\noindent with and $z=-E^{2}-i\epsilon$ and $\Lambda$ is a cutoff constant that was introduced to regulate the integral $I_{2}(z)$ by limiting $k$. It is straightforward to calculate $g\Phi E\psi(0)$ from Eq. (\ref{11})

\begin{equation}
\label{ }
	g\Phi E\psi(0) = \left[\frac{1}{g\Phi E} + \frac{\mbox{log}(\Lambda^{2}/z)}{4\pi}\right]^{-1} =  \left[\frac{1}{g\Phi E} + \frac{\mbox{log}(\Lambda/E)}{2\pi} + \frac{i}{2}\right]^{-1}~.
\end{equation}

To obtain the scattering amplitudes, we write the wave functions in position space

\begin{equation}
\label{ }
	\psi(\vec{r}) = \mbox{e}^{i\vec{k}\cdot\vec{r}} - g\Phi E\psi(0)G_{k}(r)~,
\end{equation}

\noindent where $G_{k}(r)$ is Green's function in two dimensions

\begin{equation}
\label{ }
	(-\nabla^{2} - k^{2})G_{k}(r) = \delta(\vec{r})~,
\end{equation}

\begin{equation}
\label{green}
	G_{k}(r) = \frac{i}{4}H_{0}^{(1)}(kr) \stackrel{r \rightarrow \infty}{\longrightarrow}  \frac{1}{2\sqrt{2\pi kr}}\mbox{e}^{i(kr + \pi/4)}~.
\end{equation}

\noindent By identifying the scattering amplitude from the asymptotic behavior of the scattering wave function

\begin{equation}
\label{ }
	\psi(\vec{r}) \rightarrow \mbox{e}^{i\vec{k}\cdot\vec{r}} +\frac{1}{\sqrt{r}}f(\theta)\mbox{e}^{i(kr + \pi/4)}~,
\end{equation}

\noindent we get for the \textit{constant} scattering amplitude 

\begin{equation}
\label{ }
	f(\theta) = -\frac{1}{\sqrt{2\pi E}}\frac{g\Phi E}{2}\psi(0)~,
\end{equation}

\noindent since $k^{2}=E^{2}$. Since there is no dependence on the scattering angle in $f(\theta)$ the scattering from a $\delta$ function is completely isotropic. The total cross-section in 2 dimensions is given by $\sigma_{tot} = \int_{0}^{2\pi} |f(\theta)|^{2} d\theta$. Hence, by expanding $f(\theta)$ to first order in $g$ we find that

\begin{equation}
\label{2ds}
	\sigma^{\delta}_{tot} = \frac{g^{2}\Phi^{2}E}{4}~.
\end{equation}

Our primary motivation comes from the QCD inspired axions, with mass up to the $\sim$1 eV range. To estimate the magnitude of the total cross-section, we take the value of the coupling constant $g$ from the recent result of the CAST collaboration. CAST is searching for axions produced in the sun and travelling to earth by trying to detect photons from the conversion of axions inside a constant magnetic field, following the coherent inverse Primakoff-effect. Along with the Japanese axion helioscope Sumico \cite{sumico}, CAST has set an upper limit on the magnitude of the axion-photon coupling constant of $g \lesssim 2.2 \times 10^{-10} ~\mbox{GeV}^{-1}$ for an axion mass of $m_{a} \lesssim 0.4 ~\mbox{eV}$ \cite{g}. We choose to use $g= 10^{-10}~\mbox{GeV}^{-1}$ throughout this paper. The dimensionless magnetic flux is, of course, given by $\Phi = \pi B R^{2}$, where $B$ is the magnetic field strength inside the solenoid and $R$ is the solenoid radius. Lastly, the mean energy of axions arriving at the earth from the sun is estimated to be $E = 4.2 \times 10^{3}~ \mbox{eV}$ \cite{bibber}. 

In order to get the 3D total cross-section (i.e the scattering cross-section) $\sigma_{S}$ we multiply the 2D cross-section $\sigma_{tot}$ by the length of the solenoid $L$, taking $ L = 10~ \mbox{cm}$ as an example. Multiplying the scattering cross-section by the flux of axions coming from the sun, $F = 3.67\times 10^{11} ~/\mbox{cm}^{2}\cdot\mbox{sec}$ \cite{bibber}, we can estimate the number of events per second $N$.

The quantity we are ultimately looking for is the axion-photon conversion probability. To obtain this, we calculate the ratio between the number of axions arriving at the solenoid to the number of photons produced. The number of axions hitting the solenoid is given by multiplying the flux of axions arriving by the geometrical cross section of the solenoid, given by $\sigma_{G} = DL$, where $D$ is the solenoid diameter and $L$ is its length. The number of produced photons is found by multiplying the scattering cross section ($\sigma_{S} = \sigma_{tot}\cdot L$) times the flux. Thus, the probability is given by 

\begin{equation}
\label{ }
	P_{\delta} =\sigma_{S}/\sigma_{G} = \frac{g^{2}\Phi^{2}E}{4D} = \frac{\pi^{2}g^{2}B^{2}R^{3}E}{8}~.
\end{equation} 

\noindent
Notice that the dependence on the magnetic field strength is squared. However, the dependence on the surface magnetic field gradient is ``hidden'', since it was implied  in deriving this relation. A few examples for the cross-section, number of events and probability are given below in TABLE I. 

\begin{table}[htdp]
\begin{center}\begin{tabular*}{0.65\textwidth}{@{\extracolsep{\fill}}|ccccc|}\hline 
$B$ [Tesla] & $D$ [cm] & $\sigma^{\delta}_{tot}$ [cm] & $N_{\delta}$ [$\mbox{sec}^{-1}$] & $P_{\delta} = \sigma_{S}/\sigma_{G}$ \\
10 & 1 & $3.08 \times 10^{-15}$ & $0.01$& $3.08 \times 10^{-15}$ \\
10 & 10 & $3.08 \times 10^{-11}$ & $112.98$ & $3.08 \times 10^{-12}$ \\
6 & 2 & $1.77 \times 10^{-14}$ & $0.07$ & $8.87 \times 10^{-15}$ \\
6 & 20 & $1.77 \times 10^{-10}$ & $650.78$ & $8.87 \times 10 ^{-12}$ \\
\hline \end{tabular*} 
\caption{Total cross-section, number of events and axion-photon conversion probablity for different choices of the magnetic field strength ($B$) and the solenoid diameter ($D$) and for $g= 10^{-10}~\mbox{GeV}^{-1}$. We have used rationalized natural units to convert the magnetic field units from Tesla to eV$^{2}$, where the conversion is $1~\mbox{T} = 195~ \mbox{eV}^{2}$ (please see appendix A in \cite{units} for more details).}
\end{center}
\label{table1d}
\end{table}

\section{Finite Sized Solenoidal Generated Potentials}

\subsection{Gaussian Distributed Magnetic Field}
\label{gauss}

We wish to obtain eventually measurable quantities which can be incorporated in a laboratory experiment, thus we have to consider a more realistic function to describe the magnetic field generated by the solenoid. As a first model, we choose to describe the inhomogeneous magnetic field by a Gaussian distribution around the solenoid's axis. 

\begin{equation}
\label{ }
	\vec{B}(r) = B_{0}\mbox{e}^{\frac{-r^{2}}{R^{2}}} \hat{z}~.
\end{equation}

Introducing again Green's function in the $x~,y$ plane, we write the wave function in position space

\begin{equation}
\label{psi}
	\psi(\vec{r}) = \psi_{free}(\vec{r}) + \int G(\vec{r} - \vec{\rho}) gEB(\rho)\psi(\vec{\rho})d^{2}\rho~,
\end{equation}

\noindent where $\psi_{free} = \mbox{e}^{i\vec{k}\cdot\vec{r}}$ is the solution of the free field equation. To first Born approximation, noting that  

\begin{equation}
\label{ }
	\mbox{e}^{i\vec{k}\cdot\vec{\rho}}\mbox{e}^{ik|\vec{r} - \vec{\rho}|} = \mbox{e}^{ikr}\mbox{e}^{i\left(\vec{k} - k\frac{\cdot\vec{r}}{r}\right)\cdot\vec{\rho}} ~
\end{equation}

\noindent and using again the asymptotic approximation of Green's function in 2 dimensions (see Eq. \ref{green}) we arrive at

\begin{equation}
\label{ }
	\psi(\vec{r}) = \mbox{e}^{i\vec{k}\cdot\vec{r}} + \frac{\mbox{e}^{ikr}}{2\sqrt{2\pi rE}}\int gEB(\vec{\rho})\mbox{e}^{i\vec{q}\cdot\vec{\rho}}d^{2}\rho~,
\end{equation}

\noindent where $\vec{q} = \vec{k} - k\frac{\vec{r}}{r}$. To evaluate the integral, $B(\vec{q}) = \int B(\vec{\rho})\mbox{e}^{i\vec{q}\cdot\vec{\rho}}d^{2}\rho$, we write $\vec{\rho}\cdot\vec{q} = q\rho\cos\phi$ and get

\begin{equation}
\label{ }
	B_{0}\int_{0}^{\infty}\mbox{e}^{\frac{-\rho^{2}}{R^{2}}}\rho d\rho \int_{0}^{2\pi} d\phi \mbox{e}^{iq\rho\cos\phi} = 2\pi B_{0}\int_{0}^{\infty} \mbox{e}^{\frac{-\rho^{2}}{R^{2}}}\mbox{J}_{0}(q\rho)\rho d\rho = \pi B_{0}R^{2}\mbox{e}^{-\frac{1}{4}(Rq)^{2}}~.
\end{equation}

\noindent Hence, the wave function becomes

\begin{equation}
\label{ }
	\psi(\vec{r}) = \mbox{e}^{i\vec{k}\cdot\vec{r}} + \frac{\sqrt{\pi} gB_{0}R^{2}\sqrt{E}}{2\sqrt{2 r}}\mbox{e}^{-\frac{1}{4}(Rq)^{2}}\mbox{e}^{i(kr+\pi/4)}~.
\end{equation}

\noindent By defining, as before,  

\begin{equation}
\label{ }
	\psi(\vec{r}) \rightarrow \mbox{e}^{i\vec{k}\cdot\vec{r}} + \frac{1}{\sqrt{r}}f(\theta)\mbox{e}^{i(kr + \pi/4)}~,
\end{equation}

\noindent we find for the scattering amplitude

\begin{equation}
\label{ }
	f(\theta) = \sqrt{(\pi/8)}gB_{0}R^{2}E^{1/2}\mbox{e}^{-\frac{1}{4}(Rq)^{2}}~,
\end{equation}

\noindent where the explicit dependence of $q$ on the angle is given by

\begin{equation}
\label{qs}
	q^{2} = 2k^{2}(1-\cos\theta) = 4k^{2}\sin^{2}(\theta/2)~.
\end{equation}

Hence, The total 2D cross-section is given by

\begin{equation}
\label{ }
	\int_{0}^{2\pi}|f(\theta)|^{2}d\theta = \frac{\pi}{8} (gB_{0})^{2}R^{4}E\int_{0}^{2\pi}\mbox{e}^{-\frac{1}{2}(Rq)^{2}}d\theta = \frac{\pi^{2}}{4}(gB_{0})^{2}R^{4}E\mbox{e}^{-(Rk)^{2}}I_{0}((Rk)^{2})~,
\end{equation}

\noindent where $I_{0}(x) = J_{0}(ix)$ is the modified Bessel function. The argument of this function (i.e $ (Rk)^{2}$) is very large (1 eV $\times$ 1 cm $\approx 10^{5}$) so we can use the asymptotic from of the modified Bessel function

\begin{equation}
\label{ }
	I_{n}(x) = \frac{\mbox{e}^{x}}{\sqrt{2\pi x}}\left(1 + \frac{(1-2n)(1+2n)}{8x} + ...\right)~.
\end{equation}

\noindent Keeping only the first order term gives the result 

\begin{equation}
\label{ }
	\sigma^{Gauss}_{tot} = \frac{\pi^{3/2}}{\sqrt{32}}g^{2}B_{0}^{2}R^{3} ~.
\end{equation}

Again, we find the axion-photon conversion probability $P=\sigma_{S}/\sigma_{G}$ to be

 \begin{equation}
\label{ }
	P_{Gauss} =\frac{\pi^{3/2}}{8\sqrt{2}}g^{2}B_{0}^{2}R^{2}~. 
\end{equation}
 
\noindent a result which is about two times larger than the 1D case \cite{gbl} (when taking the linear dimension associated with the extent of the magnetic field as the solenoid's radius).

\begin{table}[htdp]
\begin{center}
\begin{tabular*}{0.72\textwidth}{@{\extracolsep{\fill}}|ccccc|}\hline 
$B ~[\mbox{Tesla}]$ & $D ~[\mbox{cm}]$ & $\sigma^{Gauss}_{tot} ~[\mbox{cm}]$ & $N_{Gauss} ~[\mbox{sec}^{-1}]$ & $P_{Gauss}$ \\
10 & 1 & $1.17 \times 10^{-23}$ & $4.29 \times 10^{-11}$ & $1.17 \times 10 ^{-23}$ \\
10 & 10 & $1.17 \times 10^{-20}$ & $4.29 \times 10^{-8}$ & $1.17 \times 10^{-21}$ \\
6 & 2 & $3.38 \times 10^{-23}$ & $1.24 \times 10^{-10}$ & $1.69 \times 10^{-23}$ \\
6 & 20 & $3.38 \times 10^{-20}$ & $1.24 \times 10^{-7}$ & $1.69 \times 10^{-21}$ \\
\hline \end{tabular*}
\caption{Total 2D cross-section, number of events and the axion-photon conversion probablity for different choices of the magnetic field strength ($B$) and the solenoid diameter ($D$) for the finite sized solenoid with Gaussian distributed magnetic field case. Again, we use $g= 10^{-10}~\mbox{GeV}^{-1}$ and rationalized natural units to convert the magnetic field units from Tesla to eV$^{2}$, where the conversion is $1~\mbox{T} = 195~ \mbox{eV}^{2}$.}
\end{center}
\label{tablegauss}
\end{table}

\subsection{Solenoidal Generated Potential - Square Well Approximation}
\label{step}

Now we turn to consider the magnetic field generated by an ideal solenoidal current which is described by a step function realizing a uniform magnetic field pointing in the $\hat{z}$ direction and constrained to a cylindrical region around the origin

\begin{equation}
\label{ }
	\vec{B}(r) = \begin{cases} 
			B_{0} \hat{z}~,& r < R~,\cr 
			0~,& r > R~.
		       \end{cases}	
\end{equation}

\noindent Repeating the same manipulation as in equations (\ref{psi}) to (\ref{qs}) and using the Fourier transformation of the step function

\begin{equation}
\label{ }
	B_{0}\int_{0}^{R}\rho d\rho \int_{0}^{2\pi} d\phi \mbox{e}^{iq\rho\cos\phi} = 2\pi B_{0}\int_{0}^{R}\rho d\rho \mbox{J}_{0}(q\rho) = \frac{2\pi RB_{0}}{q}\mbox{J}_{1}(qR)~,
\end{equation}

\noindent we find that the scattering amplitude is now given by

\begin{equation}
\label{ }
	f(\theta) = \sqrt{\frac{\pi}{2}}\frac{B_{0}RgE^{1/2}}{q}\mbox{J}_{1}(qR)~.
\end{equation}

\noindent where the explicit dependence of $q$ on the angle is given by Eq. (\ref{qs}).

Before evaluating the integral for the total cross-section, let us write the total cross section for the square well case in terms of the delta function cross-section, calculated in section \ref{delta}

\begin{equation}
\label{inti}
	\sigma_{tot.}^{well} = \frac{\pi}{32}g^{2}B^{2}D^{4}E\left[\int_{0}^{2\pi}\left|\frac{J_{1}(qR)}{qR}\right|^{2}d\theta\right] = \sigma_{tot.}^{\delta}\frac{2}{\pi}\left[\int_{0}^{2\pi}\left|\frac{J_{1}(qR)}{qR}\right|^{2}d\theta\right] = \sigma_{tot.}^{\delta}\frac{2}{\pi}I(ER) ~,
\end{equation}

\noindent where $I(ER) = \int_{0}^{2\pi}\left|\frac{J_{1}(qR)}{qR}\right|^{2}d\theta$ is a dimensionless quantity which is a function of the multiplication $E\cdot R$. Using the relation $P = \sigma_{tot}/D$ (where we use the same notations as in section \ref{delta}), the proportionality constant connects also the conversion probabilities for the $\delta$ function and square well cases

\begin{equation}
\label{prop}
	P_{well} = P_{\delta}\frac{2}{\pi}I = \frac{\pi}{32}g^{2}B^{2}D^{3}EI(ER) ~.
\end{equation}

Denoting $ER = kR$ by $\eta$, the integral can be analytically solved with the solution

\begin{equation}
\label{ }
	I(\eta) = \frac{\pi}{2} ~_{2}F_{3}(\{ \tfrac{1}{2},\tfrac{2}{3} \}; \left\{ 1,2,3 \right \}; -4\eta^{2}  )~,
\end{equation} 

\noindent where $_{2}F_{3}$ is an hypergeometric function. 

To analyze this solution we expand the hypergeometric function $_{2}F_{3}$ to a series. Then, for small $\eta$, $I(\eta)$ is converging toward the constant value $\pi/2$, thus giving the equality $\sigma_{tot.}^{well} = \sigma_{tot.}^{\delta}$. This result is expected since considering only small $\eta$ values is equivalent to considering isotropic scattering because  $\eta\ll 1$ means that $ER\ll 1$. Hence, the wavelength of $\Psi$ is very large compared to the length scale of the potential. Therefore, this approximation corresponds to $\delta$ function limit of the step function, which, in turn, means that we consider isotropic scattering. 

This conclusion can also be deduced from the following viewpoint regarding the scattering angle: The integrand of $I$ is becoming extremely oscillatory as its argument (i.e. $qR$) is bigger and therefore for a reasonable scale of $ER~ (\approx 10^{5})$ we have a highly oscillatory integrand which is also decaying very fast as a function of $\theta$ (since the momentum transfer $q$ is a function of the scattering angle). Thus, the biggest contribution will come from smaller angles. In fact, demanding that the integrand will be of order one is equivalent to considering scattering angles that satisfy $\theta \lesssim 1/ER \approx 10^{-5}$. Then, using the asymptotic form of the Bessel function for small arguments we have $I \approx \pi/2$ which simply gives the solution $\sigma_{tot.}^{well} = \sigma_{tot.}^{\delta} = \pi^{2}g^{2}B^{2}R^{4}E/4$. Considering only small angles is equivalent to demanding that the argument of the Bessel function will satisfy the condition $ER\cdot\mbox{sin}(\theta/2) \ll 1$. Without limiting the range of the scattering angle, this of course means that $\eta = ER \ll 1$, the condition which coincides with the small $\eta$ expansion of $I(\eta)$.

\begin{figure}[htp]
	\begin{center}
	\includegraphics[scale=0.6]{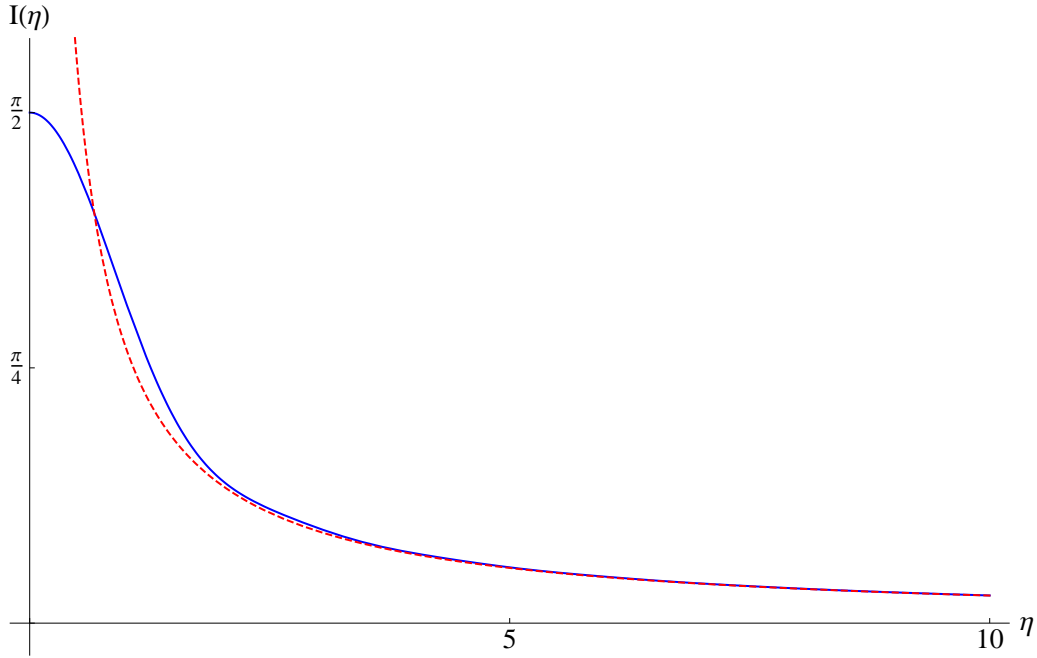}
	\end{center}
	\caption{The solution of the integral $I(\eta)$, defined by Eq. \ref{inti}, is an hypergeometric function $I(\eta) = \frac{\pi}{2} ~_{2}F_{3}(\{ \tfrac{1}{2},\tfrac{2}{3} \}; \left\{ 1,2,3 \right \}; -4\eta^{2} )$. This figure shows a plot of $I(\eta)$ as a function of the multiplication $ER = \eta$ in the solid line. The dashed line represents the fast approached limit of $I(\eta)$, which is given by $\tfrac{8}{3\pi\eta}$. At $\eta = 10$ both lines are close enough so that the ratio $\tfrac{8}{30\pi}/I(\eta=10)$ already equals $0.997$.} \label{figi}
\end{figure}

On the other end, we have the expansion for large $\eta$. This reveals the fact that the integral approaches the limit $I \rightarrow \tfrac{8}{3\pi\eta} = \tfrac{8}{3\pi E R}$ very fast. For example, for $\eta = 10$ we already have $ \tfrac{8}{30\pi}/I(\eta=10) = 0.997$. A plot of $I(\eta)$ and its limit $\tfrac{8}{3\pi\eta}$ is shown in Fig. \ref{figi}. Putting this limit into Eq. (\ref{prop}) gives the result 

\begin{equation}
\label{ }
	P_{well} = \frac{1}{6}g^{2}B^{2}D^{2} = 2\tfrac{2}{3} P_{1D} ~,
\end{equation}

\noindent where $P_{1D} = \tfrac{1}{4}g^{2}B^{2}R^{2}$ is the 1D conversion probability \cite{bibber}. Thus, the scattering from a step function potential enhances the probability of the 1D case by a factor of 2.67. 

Since the generalized hypergeometric function is difficult to analytically work with for large arguments (large $\eta$ values), we have also calculated the total 2D cross-section numerically to verify our results for the entire spectrum of $\eta$. In order to evaluate the integral $I$ we have used the 'MATLAB' program, running the new 'quadgk' function which is using the Gauss-Kronrod quadrature and is efficient specifically for oscillatory integrands. The results of the the analytical and numerical calculations match to very high precision and both results are practically the same. In fact, when considering solar axions (i.e. $ER = \eta$ is of order $10^{8}$), the numerical calculation gave the result $P_{well} = 2.67 P_{1D}$ as well. A few examples for the cross-section, number of events and conversion probability for solar axions are given below in TABLE III.  

\begin{table}[htdp]
\begin{center}
\begin{tabular*}{0.72\textwidth}{@{\extracolsep{\fill}}|ccccc|}\hline 
$B ~[\mbox{Tesla}]$ & $D ~[\mbox{cm}]$ & $\sigma^{well}_{tot} ~[\mbox{cm}]$ & $N_{well} ~[\mbox{sec}^{-1}]$ & $P_{well}$ \\
10 & 1 & $1.58 \times 10^{-23}$ & $5.80 \times 10^{-11}$ & $1.58 \times 10 ^{-23}$ \\
10 & 10 & $1.58 \times 10^{-20}$ & $5.80 \times 10^{-8}$ & $1.58 \times 10^{-21}$ \\
6 & 2 & $4.56 \times 10^{-23}$ & $1.67 \times 10^{-10}$ & $2.28 \times 10^{-23}$ \\
6 & 20 & $4.56 \times 10^{-20}$ & $1.67 \times 10^{-7}$ & $2.28 \times 10^{-21}$ \\
\hline \end{tabular*}
\caption{Total 2D cross-section, number of events and the axion-photon conversion probablity for different choices of the magnetic field strength ($B$) and the solenoid diameter ($D$) for the finite sized ideal solenoid case. We use $g= 10^{-10}~\mbox{GeV}^{-1}$ and rationalized natural units to convert the magnetic field units from Tesla to eV$^{2}$, where the conversion is $1~\mbox{T} = 195~ \mbox{eV}^{2}$}
\end{center}
\label{tablestep}
\end{table}

When comparing the cross-sections of the Gaussian distributed magnetic field to the step function, one expects the step function cross-section to be bigger than a cross-section generated by a smooth function, in agreement with similar studies done in the context of nuclear physics models, where it has been shown that a step function potential gives a bigger scattering cross-section than a smooth potential like, for example, the Woods-Saxon Diffuse potential \cite{woods}. Our results qualitatively agree with Woods and Saxon, as can be seen by comparing Tables II and III. 

\section{Resonant Scattering For $E\sim m_{a}$}
\label{rez}

So far in this report, we have consider the axion field as a massless field in order to get the $U(1)$ symmetry between axions and photons. In fact, this symmetry holds up whenever the axion mass is equal to the (effective) photon mass inside a medium. For example, in axion helioscope experiments photons acquire an effective mass if one fills the conversion region with a suitable refractive gas. 

Of course, if we had recalculated our results with massive axions and ``massive'' photons (of equal mass to that of the axion), our conclusions will have to be modified. The term that has to be taken under consideration is an $1/\sqrt{(E^{2} - m^{2})^{1/2}}$ term which comes from the Green's function and will replace the current $1/ \sqrt{E}$ in the scattering amplitude. Thus, in the $m_{a} \sim m_{photon} \neq 0$ case, the total two dimensional cross-section (for the $\delta$ function case) would have the following energy dependence

\begin{equation}
\label{res}
	\sigma_{tot} = \frac{\pi g^{2}B^{2}R^{4}E^{2}}{4\sqrt{(E^{2} - m^{2})}} ~,
\end{equation}

\noindent and we have a resonance when $E = m$, which has, in a sense, a similar behavior to the 1D problem analyzed by Adler et. al. \cite{adler} (notice that Adler et. al. consider the conversion between a massive axion and a massless photon), where of course the resonance here appears at $m_{axion} \sim m_{photon}$. In Eq. (\ref{res}) the relation $m_{axion} \sim m_{photon}$ is assumed from the beginning and we see that the additional resonance appears as $E \sim m$. For an axion rest mass below $\sim 1 ~ \mbox{eV}$, this can have practical consequences, for example, in laser generated axions (e.g in 'shining through the wall' experiments) when one can control the energy of the axion beam. 

We can see here that the 1D treatment of this process can not be justified since in the limit of zero momentum the scattering amplitude and the differential scattering cross-section become isotropic (i.e equal for all angles) and it is impossible to consider only one direction in the scattering. In fact, in the limit of exactly zero momentum (assuming there is a tunable laser capable of very fine accuracy to obtain $E$ very close to $m$) the amplitude of a finite potential becomes of the form to Eq. (\ref{res}). This is since taking the limit of zero momentum implies zero momentum transfer (from Eq. \ref{qs}) which means to consider only zero modes in the Fourier transform of the magnetic field. Hence, the cross-section of a finite potential becomes of the form of the modiÞed delta function potential. It is an experimental question whether such a fine tuning is possible with an existing laser, if the axion has a mass of the order of eV.

Achieving a resonance requires a material which has a zero index of refraction. The real part of the refractive index is given by 

\begin{equation}
\label{ }
	n_{R}(\omega) = 1 + K\frac{\omega_{0}-\omega}{(\omega_{0}-\omega)^{2} + \gamma^{2}},
\end{equation}

where $K = Ne^{2}f$ with $N$ being the number density of atoms, $e$ is the electron charge and $f$ transition oscillator strength, $\omega_{0}$ is the transition frequency and $\gamma$ represents dissipative interactions \cite{miloni}. Equating the latter to zero requires the condition $K^{2} > 4\gamma^{2}$. A negative and zero refractive indices are indeed possible as was experimentally observed by Shelby et al. \cite{shelby}. Let us hope that one day it will be possible to implement this in an axion detection lab experiment.

\section{Summary, Discussion and Conclusions}

In this paper we have studied the first examples of scattering which is not one dimensional and we have obtained enhanced probabilities. This effect is further increaed in the case of resonant scattering that appears when $E=m$ and corresponds to isotropic scattering (as in the $\delta$ function scattering). One should notice that allowing for two dimensional scattering is the same as allowing the possibility of axion-photon splitting which does not make sense in 1D scattering. We have studied here merely magnetic fields with a cylindrical structure. Further generalizations should include the scattering from, for example, a quadrupole magnetic field, which is more complicated than the cylindrical symmetric case we have studied here but, on the other hand, is quite accessible as a possible experimental setup. 

In the 1D case the conversion probability is $P_{1D} = g^{2}B^{2}l^{2}/4$ \cite{gbl}, where $l$ is the linear dimension associated with the extent of the magnetic field ($P_{1D} = P_{Gauss.}/(4\pi^{3/2})$). The comparison between the 1D scattering, finite sized potential scattering (for example, given in the table is the step-function potential) and isotropic scattering ($\delta$ function case) is given below in TABLE III. Looking at this table, we see that the effect of considering 2D scattering instead of 1D scattering increases the probability for an axion to be converted into a photon for axions coming from the sun. In fact, we see that $P_{well} = 2.67\times P_{1D}$ (as was shown in Sec. \ref{step}).

\begin{table}[htdp]
\begin{center}
\begin{tabular*}{0.81\textwidth}{@{\extracolsep{\fill}}|ccccc|}\hline 
$B~ [\mbox{Tesla}]$ & $D~ [\mbox{cm}]$ & $P_{1D}$ & $P_{\delta}$& $P_{well}$ \\
10 & 1 & $5.94 \times 10^{-24}$ & $3.08 \times 10^{-15}$ & $1.58 \times 10^{-23}$ \\
10 & 10 & $5.94 \times 10^{-22}$ & $3.08 \times 10^{-12}$ & $1.58 \times 10^{-21}$ \\
6 & 2 & $8.56 \times 10 ^{-24}$ & $8.87 \times 10 ^{-15}$ & $2.28 \times 10^{-23}$ \\
6 & 20 & $8.56 \times 10 ^{-22}$ & $8.87 \times 10^{-12}$ & $2.28 \times 10^{-21}$ \\
\hline 
\end{tabular*} 
\caption{A comparison of the axion-photon conversion probability of the 1D axion helioscope case ($P_{1D}$), the 2D delta function case ($P_{\delta}$, calculated in section \ref{delta}) and the 2D step function ($P_{2D}$, calculated in section \ref{step}). We use, as in the rest of the paper, $g= 10^{-10}~\mbox{GeV}^{-1}$ and rationalized natural units to convert the magnetic field units from Tesla to eV$^{2}$, where the conversion is $1~\mbox{T} = 195~ \mbox{eV}^{2}$}
\end{center}
\label{tablecomp}
\end{table}

When considering scattering from a finite sized potential (Gaussian and step function potentials) the enhancement of the conversion probability compared to the 1D case still gives probabilities in the same order of magnitude. This is due  to the fact that the wavelength ($1/E$) of the $\Psi$ wave function is much smaller than the length scale of the potential ($R$), which essentially results in a quasi-1D behavior of the system. When the wavelength will be smaller, or even comparable to the length scale of the potential we see that we get bigger enhancement since in this case the scattering becomes more and more isotropic and we essentially obtain $\delta$ function scattering. 

The wavelength is determined by the momentum of the particles (from the de Broglie relation). Hence, the smaller the momentum the bigger the wavelength. For the massive case, the momentum approaches zero when the energy of the particles is of the order of the particle's mass. This situation, where the wavelength of the particles is much larger than any other length scale in the problem, is realized in the resonant scattering case, discussed in Sec. \ref{rez}. There we have shown explicitly that this limit gives an isotropic scattering for a finite potential and thus, conversion probabilities of the order of the $\delta$ function case (shown in TABLE I). 

The cross-section in the resonance case was calculated at tree level. This gives a singularity of the cross-section at $E=m$. However, in practice a resonance effect should have a certain width and this, of course, should also be the case for the resonance case found here. We notice also that the resonance behavior comes together with a breakdown of the 1D treatment of axion-photon conversion and also that a finite width can be originated from absorption effects. All these problems will be addressed in a future publication.


Our results might also be applicable for the solar scenario as well. In the sun, magnetic flux tubes can play the role of a solenoidal potential while the energy spectrum of photons is continuous. Thus, we expect to have both isotropic (resonance) and anisotropic scattering. These magnetic flux tubes are enormous regions of constant magnetic flux with length scale of the order of about $10^{2}$ km in diameter and $10^{4}$ km in length. If we trust our numerical results to work in these scales as well, the conversion probability (which of course relates to at-least 2D scattering) will be greatly enhanced. For example, taking a flux tube with magnetic field of $B = 0.2$ T and diameter of $100$ km, we get a conversion probability of $P_{sun} = 6.34 \times 10^{-13}$ (with $g = 10^{-10} ~\mbox{GeV}^{-1}$), which is even larger than the isotropic scattering from a laboratory fictitious infinitely thin solenoid. 

This result may be related directly to some basic solar physics problems and, in particular, the coronal heating problem in the sun. The sun's outer layer, the solar corona, is much hotter than the surfaces below it, the chromosphere and the optical surface of the sun (the photosphere). Within a few hundred kilometers, the temperature in the corona rises to be about 500 times that of the underlying chromosphere, instead of continuing to fall to the temperature of empty space (2.7 K). While the energy flux of extreme ultraviolet photons and X-rays from the higher layers of the sun is some five orders of magnitude less than the energy flux from the photosphere, it is still surprisingly high and inconsistent with the spectrum from a black body with the temperature of the photosphere. Thus, some exotic physics must be at work out there. We would like to point out here that even the rather modest probability enhancements derived in this work, of as much as a factor of 2.67, might still provide a potential explanation we are looking for, as for why the solar X-ray activity correlates preferentially not only to magnetic fields but even more so in places with magnetic field gradient (near the inversion line of two oppositely directed magnetic field regions) \cite{zioutas}.

Moreover, as was shown earlier in this work, when the axion mass and the energy corresponding to the plasma frequency are equal the conversion probability features a resonance and  increase sharply. Since in the restless magnetic sun the magnetic fields and plasma densities are continuously changing, this resonance crossing is quite probable and can result in an otherwise unexpected photon excess or deficit. Hence, the work presented here might suggest, in its astrophysical extension, a possible solution to some basic problems in solar physics. These issues will continue to be studied by us in the future.

\vskip.6in

\centerline{{\bf Acknowledgments}} The authors wish to thank Y. Etzioni, I. Israel and A. Sadeh for their help and advice with the use of MATLAB program and to J. Jaeckel and J. Redondo for useful correspondence and conversations. We also warmly thank Y. Band for advices and useful conversions about negative and zero refractive index. E.I. Guendelman thanks the university of Trieste and INFN for the kind hospitality and support during his visit to Trieste. I. Shilon thanks the organizers of the 5th Patras workshop, and in particular J. Jaeckel, for inviting and supporting his participation in the workshop.

\vskip.3in


\end{document}